\RequirePackage{booktabs}
\documentclass[pdflatex,sn-mathphys-num]{sn-jnl}

\usepackage{graphicx}%
\usepackage{multirow}%
\usepackage{amsmath,amssymb,amsfonts}%
\usepackage{amsthm}%
\usepackage{mathrsfs}%
\usepackage[title]{appendix}%
\usepackage{xcolor}%
\usepackage{textcomp}%
\usepackage{manyfoot}%
\usepackage{booktabs}%
\usepackage{algorithm}%
\usepackage{algorithmicx}%
\usepackage{algpseudocode}%
\usepackage{listings}%

\usepackage{array}
\usepackage{hyperref}
\usepackage{tcolorbox}
\tcbuselibrary{skins, breakable, theorems}
\usepackage{threeparttable}

\usepackage[nolist]{acronym}
\begin{acronym}
  \acro{AI}{Artificial Intelligence}
  \acro{SE}{Software Engineering}
  \acro{ML}{Machine Learning}
  \acro{DL}{Deep Learning}
  \acro{FM}{Foundation Model}
  \acro{GPT}{Generative Pre-trained Transformer}
  \acro{ASE}{Automated Software Engineering}
  \acro{PTM}{Pre-trained Language Model}
  \acro{LLM}{Large Language Model}
  \acro{NLP}{Natural Language Processing}
  \acro{AI4SE}{Artificial Intelligence for Software Engineering}
  \acro{OOV}{Out-Of-Vocabulary}
  \acro{MLP}{Multi-layer Perception}
  \acro{TF-IDF}{Term Frequency–Inverse Document Frequency}
  \acro{ILF}{Inverse Location Frequency}
  \acro{CNN}{Convolutional Neural Network}
  \acro{LSTM}{Long Short-Term Memory}
  \acro{SVM}{Support Vector Machine}
  \acro{KNN}{K-Nearest Neighbor}
  \acro{DT}{Decision Tree}
  \acro{MCV}{Message Count Vector}
   
\end{acronym}

\definecolor{fbtitle}{HTML}{636463}
\definecolor{fbbg}{HTML}{F2F2F2}

\newtcbtheorem{answer}{Answer}%
  {enhanced, breakable,
    colback = fbbg, colframe = gray, colbacktitle = fbtitle,
    attach boxed title to top left = {yshift = -2.2mm, xshift = 0mm},
    boxed title style = {rounded corners},
    separator sign none,
    fonttitle = \sffamily\bfseries}{qst}

\newcommand{\etal}{\textit{et al.}}

\newcommand{\accept}[1]{}
\newcommand{\Foutse}[1]{}
\newcommand{\xingfang}[1]{}
\newcommand{\heng}[1]{}
\newcommand{\response}[2]{#2}
\newcommand{\pararesponse}[1]{#1}

\theoremstyle{thmstyleone}%
\theoremstyle{thmstyletwo}%

\theoremstyle{thmstylethree}%

\begin{document}

\title[What Information Contributes to Log-based Anomaly Detection?]{What Information Contributes to Log-based Anomaly Detection? Insights from a Configurable Transformer-Based Approach}


\author{\fnm{Xingfang} \sur{Wu}}\email{xingfang.wu@polymtl.ca}

\author*{\fnm{Heng} \sur{Li*}}\email{heng.li@polymtl.ca}

\author{\fnm{Foutse} \sur{Khomh}}\email{foutse.khomh@polymtl.ca}

\affil{\orgdiv{Department of Computer and Software Engineering}, \orgname{Polytechnique Montréal}, \orgaddress{\city{Montréal}, \state{Québec}, \country{Canada}}}


\abstract{Log data are generated from logging statements in the source code, providing insights into the execution processes of software applications and systems. State-of-the-art log-based anomaly detection approaches typically leverage deep learning models to capture the semantic or sequential information in the log data and detect anomalous runtime behaviors. However, the impacts of these different types of information are not clear. \response{R1.2 \& R3.1}{In addition, most existing approaches ignore the timestamps in log data, which can potentially provide fine-grained sequential and temporal information.} In this work, we propose a configurable Transformer-based anomaly detection model that can capture the semantic, sequential, and temporal information in the log data and allows us to configure the different types of information as the model's features. Additionally, we train and evaluate the proposed model using log sequences of different lengths, thus overcoming the constraint of existing methods that rely on fixed-length or time-windowed log sequences as inputs. With the proposed model, we conduct a series of experiments with different combinations of input features to evaluate the roles of different types of information (i.e., sequential, temporal, semantic information) in anomaly detection. The model can attain competitive and consistently stable performance compared to the baselines when presented with log sequences of varying lengths. The results indicate that the event occurrence information plays a key role in identifying anomalies, while the impact of the sequential and temporal information is not significant for anomaly detection on the studied public datasets. On the other hand, the findings also reveal the simplicity of the studied public datasets and highlight the importance of constructing new datasets that contain different types of anomalies to better evaluate the performance of anomaly detection models.}


\keywords{Anomaly Detection, Log Analysis, Log Representation, Transformer}



\maketitle

\backmatter

\section{Introduction}

\accepted{\heng{The first few paragraphs of the introduction can be made more concise: just need enough context to understand the problem (lack of understanding of the role of different types of information: semantic, sequential, timing/temporal information) and motivate the study.}}

Logging is commonly used among software developers to track the runtime status of software systems. Logs, generated through logging statements within program source code, provide insight into the execution path of code. They serve as the primary source of information for understanding system status and performance issues~\cite{he2016experience}. With logs, practitioners can diagnose system failures and analyze root causes. Initially designed for human readability, logs contain elements of natural language to some extent. As systems and applications grow increasingly complex, the volume of generated logs expands, rendering manual examination impractical and inefficient~\cite{oliner2012advances}. Researchers and developers in both academia and industry have developed various automated log analysis approaches, leveraging different types of information within log data~\cite{he2021survey}. \response{R1.3}{Despite numerous studies on log-based anomaly detection, the roles of different types of information—sequential, semantic, temporal, and event occurrence—remain unclear in a comparative sense.}

Log data are semi-structural textual data following common structures defined by developers using logging libraries. Typically, an automatic log analysis workflow contains pre-processing steps that transform the semi-structural logs into structural logs with the knowledge of the structure of the log data being processed~\cite{he2021survey}. Logs, generated by logging statements comprising both static log templates and dynamic parameters during runtime, are typically separated for further processing. As we usually do not have access to the logging statements that generate the log messages, log parsers are developed to identify dynamic fields and group the logs by their templates~\cite{zhu2019tools}.


Most of the existing log-based anomaly detection approaches work on log sequences~\cite{he2016experience, chen2021experience}. These approaches require the log data to be grouped under a specific configuration into log sequences, for which anomalies are detected. Logs generated by some systems contain certain fields (e.g., block ID in HDFS dataset) by which logs can be grouped accordingly. For the log data that do not have specific identifiers to be grouped, previous approaches usually adopt a fixed-length or fixed-time sliding window grouping. These sequences serve as basic units for log anomaly detection. Besides, some approaches (e.g., Logsy~\cite{nedelkoski2020self}) focus on identifying anomalies associated with certain log templates and, therefore, work on log events without considering the contexts offered by log sequences.

Log representation is an indispensable step in automated log analysis, which transforms textual log messages into numerical vectors~\cite{wu2023effectiveness}. In most classical approaches, Message Count Vector (MCV)~\cite{xu2009detecting, lou2010mining, he2016experience}, which counts the occurrences of log templates within a log sequence, is used to perform the transformation. In these representation techniques, sequential information within log sequences is lost. There are also some approaches that directly use the sequence of the log templates to represent a log sequence~\cite{du2017deeplog}. \response{R1.2 \& R3.1}{Besides, temporal information provided by the timestamps in logs can serve as an extra feature for identifying anomalies. In some previous works~\cite{du2017deeplog, li2020swisslog, wang2024loggt}, time intervals between the occurrences of logs in a sequence are encoded and combined with other features in a vectorized manner.}
As log messages usually contain natural language, in order to harness the semantic information therein, embedding techniques or pre-trained language models from natural language processing are employed in more recent and advanced approaches~\cite{le2021log,guo2024logformer}. Differing in their mechanisms, these methods do not necessarily require the input logs to be grouped based on\accepted{\Foutse{based on?}} their templates by the parsing process.

Machine learning models that match the obtained log representation are adopted to detect the anomalies within log events or sequences. In particular, some sequential models (e.g., CNN, LSTM, Transformer) that accept inputs of a series of log items are proposed~\cite{chen2021experience}. The utilization of sequential models is based on the intuition that log events\accepted{\heng{use consistent terms}} within sequences follow certain patterns and rules~\cite{du2017deeplog}.
Moreover, previous approaches formulate the anomaly detection task in different ways~\cite{he2016experience, chen2021experience}. Some approaches formulate the anomaly detection task as a binary classification problem and train the classifiers under a supervised scheme to classify logs or log sequences as either anomalies or normal instances\accepted{\Foutse{normal events/sequences?}}~\cite{he2016experience}. Some works formulate the task to predict future log events given sequences of past events~\cite{du2017deeplog, chen2021experience}. Once the real future events are not among the predicted candidate events, the sequence is considered to be abnormal. Moreover, some works formulate the problem by identifying pattern violations and adopting machine learning techniques, like clustering and dimension reduction, on the feature represented to find anomalies~\cite{he2018identifying,farzad2020unsupervised}.

\accepted{\heng{This paragraph is not necessary in the introduction. The goal of the introduction is just to use the least amount of context to motivate the problem and the study.}}

\accepted{\heng{This paragraph needs improvement. First, fixed-length and fixed-window are without any context. I think no need to mention them here. The paragraph can focus on the problem: existing studies consider different types of information (give examples), but it is not clear which types of information contribute to anomaly detection.}}
The existing research and approaches of log-based anomaly detection face several challenges. First, the evaluations are carried out under different settings with the same datasets, which hinders the fair comparisons of existing approaches~\cite{le2022log}. 
\response{R1.1}{There are no standardized or well-justified configurations for grouping these datasets, apart from a few limited and discrete experimental results that have explored the impact of different settings~\cite{he2016experience, wu2023effectiveness}. The uninformed choice of grouping configuration impacts the values of metrics used to evaluate anomaly detection performance and leads to a biased interpretation of the results. Moreover, when processing a new dataset, the lack of guidance on selecting an appropriate configuration can lead to suboptimal performance and unreliable conclusions.}
Second, \response{R1.4}{existing studies~\cite{landauer2024critical} have shown that commonly used datasets mostly contain anomalies that are not associated with sequential patterns,} although various sequential models are employed. The significance of utilizing sequential information in anomaly detection is unclear, \response{R1.4}{particularly when dealing with new log data.} Third, there is a limited number of datasets available for the anomaly detection task~\cite{zhu2023loghub}. The quality and characteristics of these datasets are not clear despite their wide adoption. Fourth, timestamp as a common field for all log data, which may be informative for some kinds of anomalies associated with system performance, is usually ignored in existing \response{R2.1 \& R3.1}{Transformer-based} anomaly detection approaches~\cite{he2016experience, chen2021experience}. Understanding the role of temporal information within timestamps may be beneficial for enhancing the effectiveness of anomaly detection tasks.

\noindent \textbf{Contribution} In this study, we propose a log-based anomaly detection method that is based on a Transformer model. Our proposed model is designed to be flexible and configurable: The input feature can be configured to use any combination of semantic, sequential, and temporal information, which provides flexibility for the evaluations.
\accepted{\heng{be more specific: can be configured to use any combination of semantic, sequential, and timing information}\xingfang{comments say it is repetitive with the following info.}} 
Moreover, the model can handle log sequences of different lengths, which alleviates the requirement of strict settings in the log grouping process that is commonly done in prior works~\cite{he2016experience, chen2021experience}.\accepted{\Foutse{we should explain how/if this approach addresses the problem listed above!! there is a missing connection}} We evaluate our proposed approaches against baseline approaches over four public datasets. Furthermore, we utilize the proposed model as a tool to conduct extensive evaluations that aim to strengthen the understanding of the anomaly detection task and the characteristics of public datasets commonly used to evaluate the anomaly detection methods from an experimental point of view. Specifically, we aim to examine the roles of various types of information inherent in log sequences for the anomaly detection task. For instance, by incorporating the temporal or sequential information from log sequences into their representation, we investigate the importance of utilizing this information for anomaly detection on the studied datasets. \response{R2.1}{Our proposed model could serve as a tool to analyze future datasets for the log-based anomaly detection task and arrive at conclusions that would typically require manual empirical studies and numerous experiments.}

\noindent \textbf{Research Questions}
We organize our study with the following research questions (RQs):

\textbf{RQ1:} How does our proposed anomaly detection model perform compared to the baselines?\accepted{\heng{How does our proposed anomaly detection model perform compared to the baselines?}}

\textbf{RQ2:} How much does the sequential and temporal information within log sequences affect anomaly detection?\accepted{\heng{How does the sequential and temporal information within log sequences affect anomaly detection?}}

\textbf{RQ3:} How much do the different types of information individually contribute to anomaly detection?


\noindent \textbf{Paper Organization} The rest of the paper is organized as follows. Section~\ref{two:sec:background} introduces background information about log-based anomaly detection tasks and discusses works closely related to our study.
Section~\ref{two:sec:approach} describes the design of our Transformer-based anomaly detection model used in the experiments. Section~\ref{two:sec:expsetup} details the experimental setup. We organize the experimental results by the three research questions proposed in Section~\ref{two:sec:expresults}. In Section~\ref{two:sec:discussion}, we further discuss the results and summarize the findings. Section~\ref{two:sec:threats} identifies the threats to validity of our experiments and findings. At last, we conclude our work in Section~\ref{two:sec:conclusion}.

\section{Background and Related Work}\label{two:sec:background}
\subsection{Different Formulations of the Log-based Anomaly Detection Task}
Previous works formulate the log-based anomaly detection task differently. Generally, the common formulations can be classified into the following categories.

\paragraph{\textbf{Binary Classification}} The most common way to formulate the log-based anomaly detection task is to transform it into a binary classification task where machine learning models are used to classify logs or log sequences into anomalies and normal samples~\cite{he2016experience}. Both supervised~\cite{bodik2010fingerprinting, chen2004failure, liang2007failure} and unsupervised~\cite{xu2009detecting} classifiers can be used under this formulation. In unsupervised schemes, a threshold is usually employed to determine whether it is an anomaly based on the degree of pattern violation~\cite{xu2009detecting}.

\paragraph{\textbf{Future Event Prediction}} There are also some approaches that formulate the anomaly detection task as a prediction task~\cite{du2017deeplog}. Usually, sequential models are trained to predict the potential future events given the past few logs within a fixed window frame. In the predicting phase, the models are expected to generate a prediction with Top-N probable candidates for a future event. If the real event is not among the predicted candidates, the unexpected log is considered an anomaly which violates the normal pattern of log sequences.

\paragraph{\textbf{Masked Log Prediction}} The log-based anomaly detection task can also be formulated as a masked log prediction task~\cite{guo2021logbert}, where models trained with normal log sequence data are expected to predict the randomly masked log events in a log sequence. Similar to future event prediction, a log sequence is considered normal if the actual log events that appeared in log sequences are among the predicted candidates.

\paragraph{Others} Some works formulate the anomaly detection task as a \textbf{clustering task}, where feature vectors of normal and abnormal log sequences are expected to fall into different clusters~\cite{lin2016log}. The prediction of the label for the log sequence is determined based on the distance between the sequence to be processed and the centroids of the clusters. Moreover, there are previous approaches that utilize \textbf{invariant mining}~\cite{lou2010mining} to tackle the task. They identify anomalies by discerning pattern violations of feature vectors of log sequences.

\subsection{Supervised v.s. Unsupervised}
Another dimension of the formulations of the anomaly detection tasks is based on the training mechanisms. Supervised anomaly detection methods demand labeled logs as training data to learn to discern abnormal samples from normal ones, while unsupervised methods learn from the normal pattern from normal log data and do not require labels in the model training process. Unsupervised methods offer greater practicality as we do not usually have access to well-annotated log data. However, supervised methods usually achieve superior and more stable performance according to previous empirical studies~\cite{he2016experience}.

\subsection{Information within Log Data} Generally, log data that is formed by sequences of log events contains various types of information. Within a log sequence, the \textbf{occurrences of logs} from different templates serve as a context and are a distinctive feature for log sequences. Similar to the Bag-of-Words model, numerical presentation based on the frequency of the template occurrences can represent log sequences and be used in anomaly detection. Various works~\cite{he2016experience} utilize the \acs{MCV} to represent this information. Moreover, the \textbf{sequential information} within the log items provides richer information about the occurrences of logs and probably reflects the execution path of applications and services. DeepLog~\cite{du2017deeplog} uses a \acs{LSTM} model to encode the sequential information. Furthermore, the \textbf{temporal information} from the log data provides even richer details about the occurrence of logs. The time intervals between log events may offer valuable insights into anomaly detection and other log analysis tasks about the system status, workload, and potential blocks. \response{R1.2 \& R3.1}{For example, Du~\etal.~\cite{du2017deeplog} combined time intervals with other log parameters to construct the time series input for a Parameter Value Anomaly Detection model. In Swisslog~\cite{li2020swisslog}, the time interval is encoded using soft one-hot encoding, and the resulting time embedding vector is combined with the semantic embedding as input to an Attn-Bi-LSTM architecture. The concatenation of embeddings resulted in a high dimensionality, which leads to an increased computational cost. LogGT~\cite{wang2024loggt} encodes time intervals within a graph structure, allowing it to capture temporal correlations and detect anomalies caused by temporal changes in log sequences. Although these studies incorporate time intervals, their impact on anomaly detection remains unclear due to inconsistent encoding methods and lack of systematic analysis. A more thorough evaluation is needed to assess their effectiveness.}


Besides, \textbf{textual or semantic information} provided by log messages has garnered significant attention in recent studies~\cite{chen2021experience, le2021log, guo2024logformer}. Given the inherent nature of log data, log messages written by developers articulate crucial information in natural language regarding the system's operations, errors, and events, making them valuable for troubleshooting and system analysis. Various natural language processing techniques are employed to extract textual features and generate embeddings for log messages. From basic numerical statistics such as TF-IDF to word embedding techniques like Word2Vec, and advancing to advanced contextual embedding methods like BERT, these advancements are geared towards more accurately capturing the semantic information contained within log messages. Their objective is to distinguish between unrelated logs and connect similar ones, thereby supplying more informative and distinguishable features for subsequent downstream models.

In addition, the \textbf{parameters} carried by the log messages offer more diverse information about the systems. However, as most parameters are system-specific and lack a consistent format or range, deciding on the best way to model the information from different parameters is a formidable challenge. In most previous works, the parameters that are usually numbers and tokens are removed in pre-processing stages. In DeepLog~\cite{du2017deeplog}, a parameter value anomaly detection model for each log key (i.e., log template) is used to detect anomalies associated with parameter values as an auxiliary measure to the log key anomaly detection model. In a more recent study~\cite{guo2024logformer}, a parameter encoding module is employed to produce character-level encodings for parameters. Following this, each output is assigned a learnable scalar, which functions as a bias term within the self-attention mechanism. Moreover, log data generated by various systems and applications often contains system-specific information that may require domain-specific knowledge and tailored approaches to optimize the performance of downstream tasks. \accepted{\heng{Like for the parameter information, it would be good to discuss example approaches for other types of information.}}

\response{R1.3}{Prior studies mainly examined these types of information with different feature extraction approaches individually, without systematically comparing their relative contributions to the performance of anomaly detection. Event occurrence information has been the most commonly used and widely encoded feature in previous log-based anomaly detection studies. Empirical studies~\cite{wu2023effectiveness, landauer2024critical, yu2024deep} have shown that methods leveraging event occurrences can achieve strong performance on certain datasets, yet their effectiveness has not been systematically evaluated in relation to other log features, leaving its comparative importance unclear. Sequential information is implicitly captured by modern deep learning-based architectures, such as LSTMs~\cite{du2017deeplog, li2020swisslog} and Transformers~\cite{guo2021logbert,le2021log,guo2024logformer}. However, empirical findings suggest that its effectiveness is highly dataset-dependent, as anomalies in certain datasets do not necessarily manifest through sequential patterns~\cite{landauer2024critical}. As a result, when encountering new datasets, it remains unclear to what extent sequential information contributes to anomaly detection or whether sequence-aware models are the most appropriate choice. Additionally, semantic information has been used with various embedding techniques in anomaly detection~\cite{wu2023effectiveness}; however, its significance in relation to event occurrence remains underexplored. Moreover, while some studies incorporate temporal information from log timestamps, its impact on anomaly detection remains unclear due to the lack of systematic analysis. The importance of temporal information may also vary across datasets, complicating its general contribution. Determining this empirically requires extensive experimentation, as there is no simple or direct method to assess the significance of sequential patterns in a given dataset. \textbf{This study systematically evaluates these different types of information within a configurable Transformer-based framework, providing new insights into their impact on log-based anomaly detection.}}

\subsection{Fixed-Window Grouping}
\accepted{\heng{I feel this may be moved to the end of this section as it is very detailed compared to other aspects?}}
Available public datasets for log-based anomaly detection have either sequence-level or event-level annotations. For the datasets that do not have a grouping identifier, fixed-length or fixed-time grouping is often employed in the pre-processing process to form log sequences that can be processed by log representation techniques and anomaly detection models.\accepted{\heng{Need some context to explain why log grouping is needed for event-level annotations}} Various grouping settings have been used in previous studies for public datasets~\cite{he2016experience}. The different grouping settings generate different amounts of samples and varying contextual windows of log data, making direct comparisons of their performance impossible. Moreover, the logs are not generated with fixed rates or fixed lengths. Using fixed-window grouped log sequences for training and testing samples does not align with the actual scenarios.


\response{R2.1}{\subsection{Existing Transformer-based Anomaly Detection Methods}\label{sec:TransSOTA}}

\response{R2.1}{}
\pararesponse{The Transformer~\cite{vaswani2017attention} is a deep learning architecture that relies on the self-attention mechanism to process sequential data. Due to its strong ability to capture dependencies and learn complex patterns, it has become the foundation for various sequence modeling tasks (e.g., NLP tasks). Given its advantages and the similarities between log data and NLP data, the Transformer has emerged as a promising building block for log analysis models. In particular, Transformer-based models have been explored for log anomaly detection, leveraging their ability to capture complex dependencies and contextual relationships within log sequences.}

\pararesponse{Previous works on transformer-based log anomaly detection have adopted diverse implementations, varying in architecture, feature extraction and encoding methods, and training strategies. Logsy~\cite{nedelkoski2020self} uses a Transformer encoder that extracts and aggregates contextual relationships of logs into a special [EMBEDDING] token, which is then optimized using a hyperspherical classification loss to cluster normal logs near the origin and flag anomalies by their distance. However, the method relies heavily on auxiliary anomaly data, is sensitive to hyperparameter settings, and can be computationally expensive with diverse or limited log datasets. NeuralLog~\cite{le2021log} combines semantic embeddings with positional encoding to model log sequences, where each log message is preprocessed and vectorized into an embedding with a pre-trained BERT model without the need for log parsing. It employs a straightforward preprocessing step that removes special tokens and numeric characters from log messages, thus eliminating the need for log parsing. However, because log messages are not grouped by their corresponding templates, the pre-trained BERT model redundantly encodes identical log events each time they appear, leading to unnecessary computational overhead. In contrast, LogBERT\cite{guo2021logbert} learns log key embeddings during training, but by using only the log keys extracted by a parser as input, it overlooks the richer semantic information contained in log messages. Although this approach includes ``BERT" in its name, it uses the Transformer encoder as its fundamental structure to learn and represent contextual relationships among log keys. This is achieved through two training objectives: Masked Log Key Prediction, which mimics BERT's self-supervised training scheme, and Volume of Hypersphere Minimization, which regulates the distribution of normal log sequences. Same as previously proposed methods, it does not consider the temporal relationships among logs. Instead of utilizing the standard self-attention structure, LogFormer~\cite{guo2024logformer} proposed Log-Attention module which aims to include the parameter encoding to alleviate the information loss caused by log parsing. An adapter structure is used to transfer knowledge from the source domain to the target domain in order to alleviate the poor generalization on multi-domain logs. This approach relies on a log parser to extract parameters from logs and a pre-trained BERT model to encode the log templates.}

\pararesponse{Existing Transformer-based approaches have focused on different aspects of log-based anomaly detection and dealed with specific issues of existing approaches (e.g., the generalization~\cite{nedelkoski2020self}, errors induced by log parser~\cite{le2021log}, the shortage of anomalous data~\cite{guo2021logbert}, the transferability~\cite{guo2024logformer}). Therefore, the performances of these methods are evaluated under different settings with various training schemes. Still, fixed-window grouping is still the most common method for grouping logs into sequences, which are the basic units these methods operate on. Besides, all of these methods hold the hypothesis that anomalies within logs are associated with sequential patterns. Moreover, none of them considered the timestamps in the log data, thereby neglecting the inherent temporal relationships between log entries.}


\subsection{Related Empirical Studies}\label{two:sec:related}
Recent empirical studies on log-based anomaly detection aim to deepen the understanding of the existing log-based anomaly detection models and the public datasets for evaluation. They focus on several issues. Le~\etal~\cite{le2022log} conducted an in-depth analysis of recent deep-learning anomaly detection models over several aspects of model evaluation. Their findings suggest that different settings of stages in anomaly detection would greatly impact the evaluation process. Therefore, using diverse datasets and analyzing logical relationships between logs are important for assessing log-based anomaly detection approaches.

Wu \etal~\cite{wu2023effectiveness} conducted an empirical study on vectorization (i.e., representation) techniques for log-based anomaly detection. They evaluated the effectiveness of some existing classical and semantic-based techniques with different anomaly detection models. Their experimental results suggest that the classical ways of transforming textual logs into feature vectors can achieve competitive results with more complex semantic embeddings.

A more recent work~\cite{yu2024deep} compared classical and deep-learning approaches of log-based anomaly detection methods. Their experimental results also suggest that simple models can outperform complex log vectorization methods. The deep learning approaches fail to surpass the simpler techniques. Their work highlights the need to critically analyze the datasets used in evaluation. 

Moreover, Landauer \etal~\cite{landauer2024critical} critically reviewed the common log datasets used to evaluate anomaly detection techniques. Their analysis of the log datasets suggests that most anomalies are not directly associated with sequential information within the log sequence. Sophisticated detection methods are unnecessary for attaining excellent detection performance. Their findings also highlight the creation of new datasets that incorporate sequential anomalies for evaluating anomaly detection approaches.

\textbf{In our work, we proposed a Transformer-based anomaly detection model capable of capturing sequential and temporal information within the log sequence, in addition to event occurrence and semantic information. \response{R2.1}{The proposed model can serve as a tool for characterizing datasets and anomalies in log-based anomaly detection by comparing the significance of different patterns and types of information.}} Thanks to the flexibility of the proposed model, we can easily utilize various combinations of log features as input for our evaluations. Through a series of carefully designed experiments, we scrutinized the four common public datasets and deepened our understanding of the roles of different types of information in identifying anomalies within the log sequence. Our findings are generally in accordance with the previous empirical studies. However, our analysis offers a more comprehensive and detailed understanding of the anomaly detection task and the studied public datasets.

\section{A Configurable Transformer-based Anomaly Detection Approach}\label{two:sec:approach} \accepted{\heng{Maybe title the section as ``A Configurable Transformer-based Anomaly Detection Approach''? ``Approach'' is too general and should contain how you answer the research questions.}}

In this study, we introduce a novel Transformer-based method for anomaly detection. The model takes log sequences as inputs to detect anomalies. The model employs a pre-trained BERT model to embed log templates, enabling the representation of semantic information within log messages. These embeddings, combined with positional or temporal encoding, are subsequently inputted into the Transformer model. The combined information is utilized in the subsequent generation of log sequence-level representations, facilitating the anomaly detection process. We design our model to be flexible: The input features are configurable so that we can use or conduct experiments with different feature combinations of the log data.
Additionally, the model is designed and trained to handle input log sequences of varying lengths. In this section, we introduce our problem formulation and the detailed design of our method.


\subsection{Problem Formulation}\label{two:sec:formulation}

We follow the previous works~\cite{he2016experience}\accepted{\heng{cite example papers}} to formulate the task as a binary classification task, in which we train our proposed model to classify log sequences into anomalies and normal ones in a supervised way. For the samples used in the training and evaluation of the model, we utilize a flexible grouping approach to generate log sequences of varying lengths. The details are introduced in Section~\ref{two:sec:varying_length}.


\accepted{\heng{Lack of a sub-section about the grouping of the log sequences, even though the grouping is flexible?}\xingfang{it is in \ref{two:sec:varying_length}}\heng{It could be briefly mentioned here: it's the preparation for the input of the approach}}

\subsection{Log Parsing and Log Embedding}\label{two:sec:parsing_embedding}
In our work, we transform log events into numerical vectors by encoding log templates with a pre-trained language model. To obtain the log templates, we adopt the Drain parser~\cite{he2017drain}, which is widely used and has good parsing performance on most of the public datasets~\cite{zhu2019tools}. We use a pre-trained sentence-bert model~\cite{reimers2019sentence} (i.e., all-MiniLM-L6-v2~\cite{minilmmodel}) to embed the log templates generated by the log parsing process. The pre-trained model is trained with a contrastive learning objective and achieves state-of-the-art performance on various NLP tasks\accepted{\heng{include log analysis? if so cite}}. We utilize this pre-trained model to create a representation that captures semantic information of log messages and illustrates the similarity between log templates for the downstream anomaly detection model. The output dimension of the model is 384.

\subsection{Positional \& Temporal Encoding}\label{two:sec:enc}
The original Transformer model~\cite{vaswani2017attention} adopts a positional encoding to enable the model to make use of the order of the input sequence. As the model contains no recurrence and no convolution, the models will be agnostic to the log sequence\accepted{\heng{the models will be agnostic to the log sequence?}} without the positional encoding. While some studies suggest that Transformer models without explicit positional encoding remain competitive with standard models when dealing with sequential data~\cite{irie2019language, haviv-etal-2022-Transformer}, it is important to note that any permutation of the input sequence will produce the same internal state of the model\heng{not clear what ``produce same internal state'' mean}.\accepted{\heng{Double-check this sentence}}

As sequential information or temporal information may be important indicators for anomalies within log sequences, previous works that are based on Transformer models utilize the standard positional encoding to inject the order of log events or templates in the sequence~\cite{guo2021logbert, le2021log, guo2024logformer}, aiming to detect anomalies associated with the wrong execution order. However, we noticed that in a common-used replication implementation of a Transformer-based method~\cite{chen2021experience}\accepted{\heng{cite the work}}, the positional encoding was, in fact, omitted. \accepted{\heng{how about the other works? they don't have replicaiton package?}\xingfang{Mentioned previously}} To the best of our knowledge, no existing work has encoded the temporal information based on the timestamps of logs for their anomaly detection method. The effectiveness of utilizing sequential or temporal information in the anomaly detection task is unclear.

In our proposed method, we attempt to incorporate sequential and temporal encoding into the Transformer model and explore the importance of sequential and temporal information for anomaly detection. Specifically, our proposed method has different variants utilizing the following sequential or temporal encoding techniques. The encoding is then added to the log representation, which serves as the input to the Transformer structure.

\noindent \textbf{Integrating sequential information using positional encoding}\accepted{\heng{maybe use title ``Integrating sequential information using positional encoding''?}} In Transformer models, positional encoding is utilized to capture the positional information of tokens within a sequence~\cite{vaswani2017attention}. In our work, we utilize positional encoding to convey the sequential information of log events within log sequences to the model, allowing it to consider the order of log events and capture sequential dependencies effectively. The positional encoding technique takes the position index of each log in the sequence as input and generates encoding for the position of each log event. The positional encoding is computed according to the following equations.

\begin{equation}
    \label{two:equation:pos}
    \begin{aligned}
    PE_{(pos, 2i)} &= \sin\left(\frac{pos}{10000^{2i/d_{model}}}\right)  \\
    PE_{(pos, 2i+1)} &= \cos\left(\frac{pos}{10000^{2i/d_{model}}}\right)
    \end{aligned}
\end{equation}

\noindent where $pos$ is the position index of the log in the log sequence, $i$ is the dimensional index, $d_{model}$ is the dimension of the log representation in our case.

\noindent \textbf{Integrating temporal information using timing encoding} \accepted{\heng{maybe use title ``Integrating temporal information using timing encoding''?}}
Temporal information is richer than sequential information as we can tell the time intervals between log events besides the sequential information. To encode the temporal information into the input feature, we propose the employment of the following two temporal encoding methods.

\subsubsection{Relative Time Elapse Encoding (RTEE)}\accepted{\heng{cite}\xingfang{We proposed this.}}
We propose this temporal encoding method, RTEE, which simply substitutes the position index in positional encoding with the timing of each log event. We first calculate the time elapse according to the timestamps of log events in the log sequence. Instead of using the log event sequence index as the position to sinusoidal and cosinusoidal equations, we use the relative time elapse to the first log event in the log sequence to substitute the position index. Table~\ref{two:table:temporal_example} shows an example of time intervals in a log sequence. In the example, we have a log sequence containing 7 events with a time span of 7 seconds. The elapsed time from the first event to each event in the sequence is utilized to calculate the time encoding for the corresponding events. Similar to positional encoding, the encoding is calculated with the above-mentioned equations~\ref{two:equation:pos}\accepted{\heng{use equation number and cite the number}}, and the encoding will not update during the training process.

\begin{table}
\centering
\caption{An example showcasing the relative time intervals in a log sequence.\accepted{\heng{``Time intervals'' could be changed to ``Elapsed time'' or ``Relative time''}}}
\label{two:table:temporal_example}
\begin{tabular}{lccc}
                             & Log Events & Timestamps & Elapsed time  \\ 
\cline{2-4}
\multirow{7}{*}{\Huge\textbf{↓}} & T6         & 1131060239 & 0               \\
                             & T3         & 1131060240 & 1               \\
                             & T2         & 1131060240 & 1               \\
                             & T1         & 1131060241 & 2               \\
                             & T1         & 1131060243 & 4               \\
                             & T4         & 1131060244 & 5               \\
                             & T5         & 1131060245 & 6              
\end{tabular}
\end{table}

\subsubsection{Time2Vec Encoding} Time2Vec Encoding~\cite{kazemi2019time2vec} differs from the RTEE encoding method by its trainable property. We utilize Time2Vec to encode the time intervals within log sequences. The encoding is a vector of size $k+1$, which in our case is equal to the dimension of the log representation, defined by the following equations:

\begin{equation*}\label{two:equation:time2vec}
    \mathbf{t} 2\mathbf{v}(\tau)[i]=\left\{\begin{array}{ll}
    \omega_{i} \tau+\varphi_{i}, & \text { if } i=0. \\
    \mathcal{F}\left(\omega_{i} \tau+\varphi_{i}\right), & \text { if } 1 \leq i \leq k.
    \end{array}\right.
\end{equation*}

\vspace{2mm}

\noindent where $\mathbf{t} 2\mathbf{v}(\tau)[i]$ is the $i^{th}$ element of Time2Vec encoding for the moment of $\tau$, $\mathcal{F}$ is a periodic activation function, and $\omega_{i}$ and $\varphi_{i}$ are learnable parameters. In our work, we use the sine function as the periodic activation function $\mathcal{F}$. Same as in RTEE\accepted{\heng{RTEE?}} , we use the elapsed time relative to the first event as the input (i.e., the moment $\tau$) to the encoding functions.

\subsection{Model Structure}
\begin{figure}[!ht]
    \centering
	\includegraphics[width=0.8\textwidth]{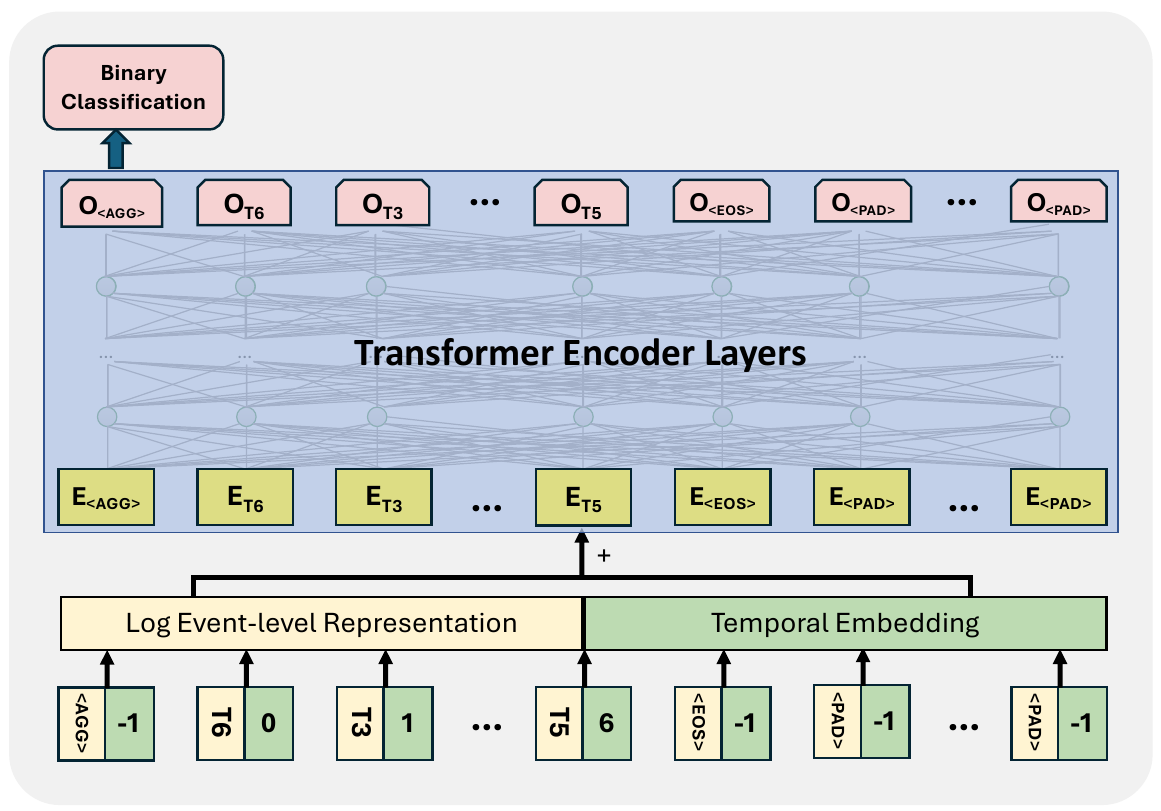}
	\caption[An illustration of our proposed Transformer-based anomaly detection model]{The illustration of the Transformer-based anomaly detection model. The figure is based on the case when the temporal encoding is used. The numbers in green cells are the time elapsed since the first log event in the sequence. We assign a value of -1 to all padded items.\accepted{ \heng{Also indicate the positional/sequential encoding (it could be an explanation, otherwise it may cause confusion). Figure can be bigger (text is too small).}} }
	\label{two:figure:Transformer}
\end{figure}

The Transformer is a neural network architecture that relies on the self-attention mechanism to capture the relationship between input elements in a sequence. The Transformer-based models and frameworks have been used in the anomaly detection task by many previous works~\cite{nedelkoski2020self,guo2021logbert, le2021log,guo2024logformer}. Inspired by the previous works, we use a Transformer encoder-based model for anomaly detection. We design our approach to \accepted{\heng{design our approach to accept ...?}}accept log sequences of varying lengths and generate sequence-level representations. To achieve this, we have employed some specific tokens in the input log sequence for the model to generate sequence representation and identify the padded tokens\heng{provide a brief rationale for the padded tokens} and the end of the log sequence, drawing inspiration from the design of the BERT model~\cite{devlin2018bert}. In the input log sequence, we used the following tokens: \textit{$<$AGG$>$} is placed at the start of each sequence to allow the model to generate aggregated information for the entire sequence, \textit{$<$EOS$>$} is added at the end of the sequence to signify its completion,
and \textit{$<$PAD$>$} is used for padded tokens. The embeddings for these special tokens are generated randomly based on the dimension of the log representation used. An example is shown in Figure~\ref{two:figure:Transformer}, the time elapsed for \textit{$<$AGG$>$}, \textit{$<$EOS$>$} and \textit{$<$PAD$>$} are set to -1. The log event-level representation \response{R2.3}{is then passed to a fully-connected layer, and the output is summed with the positional or temporal embedding, which is then fed into the encoder layers. An attention mask is passed to the self-attention mechanism in the Transformer to prevent the model from attending to padding tokens, ensuring that it focuses only on the meaningful parts of the log sequence.}


\subsection{Supervised Binary Classification} Under this training objective, we utilize the output of the first token \textit{$<$AGG$>$} of the Transformer model while ignoring the outputs of the other tokens\accepted{\heng{use ... for what...}}. This output of the first token is designed to aggregate the information of the whole input log sequence, similar to the $<$CLS$>$ token of the BERT model, which provides an aggregated representation of the token sequence.\accepted{\heng{, similar to the $<$CLS$>$ token of the BERT model which provides an aggregated representation of the token sequence?}} Therefore, we consider the output of this token as a sequence-level representation. We train the model with a binary classification objective (i.e., Binary Cross Entropy Loss) with this representation.

\section{Experimental Setup}~\label{two:sec:expsetup}
\vspace{-3mm}
\subsection{Datasets}
We evaluate our proposed approach and conduct experiments with four commonly-used public datasets: HDFS~\cite{xu2009detecting}, Blue Gene/L (BGL), Spirit, and Thunderbird~\cite{oliner2007supercomputers}. These datasets are commonly used in existing studies~\cite{he2016experience, chen2021experience, guo2024logformer}.\accepted{\heng{Mention these datasets are commonly used in existing studies (cite)}}

The HDFS dataset~\cite{xu2009detecting} is derived from the Amazon EC2 platform. The dataset comprises over 11 million log events, each linked to a block ID. This block ID allows us to partition the log data into sessions. The annotations are block-wise: each session is labeled as either normal or abnormal. In total, there are 575,061 log sessions, with 16,838 (2.9\%) identified as anomalies. The BGL, Spirit, and Thunderbird datasets are recorded from supercomputer systems, from which they are named. Different from the HDFS dataset, all these datasets have log item-wise annotation. However, there is no block ID or other identifier to group the log items into sequences. The BGL dataset is recorded with a time span of 215 days, containing 4,747,963 log items, where 348,460 (7.3\%) are labeled as anomalies. As the Spirit and Thunderbird datasets each contain more than 200 million log items, which is too large to process, we use subsets of 5 million and 10 million log items, respectively, as per the practices of previous works~\cite{le2021log, le2022log, wu2023effectiveness}. We split the datasets into an 80\% training set and a 20\% test set. For the HDFS dataset, we randomly shuffle the sessions to perform dataset splitting. \response{R1.8}{For the remaining datasets, we divide them in accordance with the chronological order of logs, as this better reflects practical usage scenarios where models are trained on past data and used to predict future events~\cite{lyu2021empirical}.} The summarised properties of datasets utilized in the evaluation and experiment of our study are presented in Table~\ref{two:table:dataset}.

\begin{table}[h!]
\centering
\caption{Characteristics of the datasets used in this study}
\label{two:table:dataset}
\begin{tabular}{cccccc} 
\toprule
Dataset     & Source             & \begin{tabular}[c]{@{}c@{}}Grouping\\Identifier\end{tabular} & \begin{tabular}[c]{@{}c@{}}Label\\Granularity\end{tabular} & \# Messages & \% Anomalies  \\ 
\hline
HDFS        & Distributed system & Block ID                                                     & Block                                                      & 11,175,629  & 2.9\%         \\
BGL         & Supercomputer      & None                                                         & Item                                                       & 4,747,963   & 7.3\%         \\
Spirit      & Supercomputer      & None                                                         & Item                                                       & 5,000,000   & 15.5\%        \\
Thunderbird & Supercomputer      & None                                                         & Item                                                       & 10,000,000  & 4.1\%         \\

\toprule

\end{tabular}
\end{table}

\subsection{Evaluation Metrics}
To measure the performance of the model in anomaly detection, we use the Precision, Recall, Specificity, and F1-score, which are calculated as follows: $Precision = \frac{TP}{TP+FP}$, $Recall = \frac{TP}{TP+FN}$, $Specificity = \frac{TN}{TN+FP}$\accepted{\heng{this metric may not be important in this context}}, $F1 = \frac{2PrecisionRecall}{Precision+Recall} = \frac{2TP}{2TP+FP+FN}$. Precision, Recall, and F1-score are commonly used to evaluate the performance of anomaly detection models. Specificity measures the percentage of normal log sequences that are correctly predicted.

\subsection{Generating Log Sequences of Varying Lengths}~\label{two:sec:varying_length}
Except for the HDFS dataset, which has a block ID to group the logs into sequences, other datasets employed by our study have no identifier to group or split the whole log sequence into sub-sequences. \response{R1.1}{In practice, the logs produced by systems and applications do not adhere to a fixed rate of generation. Therefore, logs are collected at varying rates, and users may provide chunks of logs with arbitrary lengths or spanning different time durations. Using fixed-window or fixed-time grouping with a sliding window fails to adequately accommodate the variability in log generation and thus may lead to inaccurate detection of anomalies in real scenarios.} Moreover, according to previous studies~\cite{he2016experience, le2022log, wu2023effectiveness}, the best grouping setting varies depending on the dataset, and these settings can significantly influence the performance of the anomaly detection model, making it challenging to evaluate and compare the effectiveness of anomaly detection methods.~\response{R1.1}{Therefore, evaluating anomaly models with input log sequences of varying lengths allows for a fairer comparison among different anomaly detection approaches and their settings. It also takes into account their ability to work under different context windows, making the evaluation more aligned with practical use without requiring prior knowledge of the best splitting configurations.}

In this study, we use a method to generate log sequences with varying lengths and utilize these sequences to train the model within our anomaly detection framework. In the process of log sequence generation, we determined specific parameters, including minimum and maximum sequence lengths, as well as a designated step size.\accepted{\heng{Mention that these parameters are only for the sake of evaluation in this study; in a real anomaly detection scenario, the approach can take any given sequence, right?}\xingfang{in real case, these values should be set as well.}} The step size is used to control the interval of the first log events in log sequences. The length of each log sequence is randomly generated in the range of the minimum and the maximum length. We assume the log sequence of the minimum length can offer a minimum context for a possible anomaly. The step size controls the overlaps of sequences. \response{R1.7}{The maximum length affects the input sequence length for the model and consequently influences the number of parameters, while the step size determines the number of samples in the dataset.} They should be aligned with the data distribution and computational resources available. In the experiments conducted in this study, we set the minimum length as 128, the maximum length as 512, and the step size as 64 for the datasets without a grouping identifier. 
\accepted{\heng{May need to discuss the cases with a smaller or much larger length}}

\subsection{Implementation Details and Experimental Environment}
In our experiments, the proposed Transformer-based anomaly detection model has two layers of the Transformer encoder. \accepted{\heng{Provide rationale for the choices of the hyperparameter values}}The number of attention heads is 12, and the dimension of the feedforward network layer within each Transformer block is set to 2048.~\response{R2.3}{The dimension of the fully connected layer after the log embedding layer is empirically set to 64, as experiments across a wide range of values showed that it consistently achieves good performance across all feature and dataset combinations.} We use AdamW with an initial learning rate of 5e-4 as the optimization algorithm and employ the OneCycleLR learning rate scheduler to enable a better convergence. We selected these hyperparameters following standard practices while also considering computational efficiency. Our implementation is based on Python 3.11 and PyTorch 2.2.1.
All the experiments are conducted on a high-performance computing (HPC) system. We utilize multiple computational nodes with different hardware configurations, including Intel Gold 6148 Skylake @ 2.4 GHz CPUs and NVIDIA V100 GPUs. In some cases, we use GPU instances rather than the entire GPU.

\section{Experimental Results}~\label{two:sec:expresults}

We organize this section by our research questions (RQs).


\subsection{RQ1: How does our proposed anomaly detection model perform compared to the baselines?} 

\subsubsection{Objective} This research question aims to assess the effectiveness of our proposed anomaly detection framework by evaluating its performance. We compare it against three baseline models. This research question also serves as a sanity test to ensure the validity of the results and findings of the subsequent RQs. As our objective is to check if our model works properly, we do not expect the model to outperform all the state-of-the-art methods, but it should be able to achieve competitive performance.

\subsubsection{Methodology} \label{sec:rq1:method}
\response{R1.8}{To answer this research question, we conduct two sets of experiments to compare our method with both semantic and non-semantic counterparts.}
According to a recent study~\cite{yu2024deep}, simple models can outperform complex \acs{DL} ones in anomaly detection in both time efficiency and accuracy. The superior results achieved by sophisticated \acs{DL} methods turn out to be misleading due to weak baseline comparisons and ineffective hyperparameter tuning.

Hence, in the first set of our experiments, we first employ simple machine learning methods, including \ac{KNN}, \ac{DT}, and \ac{MLP} as baselines as in Yu \etal's work~\cite{yu2024deep}. 
\response{R1.8}{We use the \acs{MCV}, which does not leverage the semantic information within log messages, as the input feature for these baseline models. We additionally conduct a grid search on the hyperparameters to strive for optimal performance for the baseline models. We compare the variant of our proposed model that utilizes a random embedding method and Relative Time Elapse Encoding (RTEE) with these baseline models. The random embedding method generates a unique feature vector for each log event, meaning the model can only detect anomalies based on the occurrence of log events in sequences, without leveraging semantic information. Consequently, the model has no access to anomalous information that may be conveyed by log message semantics. Since the inclusion of semantic information may favor our model, this comparison ensures fairness to non-semantic baselines, which might otherwise be disadvantaged due to concept drift in the log data~\cite{le2022log}. We evaluate all models using log sequences of varying lengths (see Section~\ref{two:sec:varying_length}) to assess their effectiveness and stability across different context window sizes.}

\response{R1.8}{To verify the correctness of our implementation and assess whether our proposed method achieves competitive performance against other Transformer-based approaches, we include NeuralLog~\cite{le2021log} as an additional baseline. Since NeuralLog shares a similar architecture, follows the same problem formulation, and has demonstrated strong performance, its inclusion allows us to validate our implementation and ensure that our method performs at a comparable level. Specifically, we compare the variant of our model that incorporates semantic embedding and Relative Time Elapse Encoding (RTEE) with NeuralLog.} 

Both sets of experiments are conducted on all four datasets. For the HDFS dataset, we use the Block ID to group the logs into sessions in both sets of experiments. \response{R1.8}{For datasets without a grouping identifier, the samples in the training and test sets have varying lengths, generated according to the process explained in Section~\ref{two:sec:varying_length} for the first set of experiments. As NeuralLog is designed to accept fixed-length sequences, to enable fair comparison, we use a fixed-length window grouping with a window size of 128 and a step of 64 for evaluating both models in the second set of experiments.}

\subsubsection{Result}
\response{R1.8}{}
\paragraph{Experiment Set 1: Performance Comparison with Non-Semantic Models} Table~\ref{two:table:supervised} presents the performance of our model variant and the traditional non-semantic baseline models. Overall, the variant that utilizes random embedding and RTEE achieves excellent performance on the studied datasets, with an F1 score ranging from 0.915 to 0.993. The results show that our model variant, without utilizing semantic information, achieves competitive performance compared to the non-semantic baseline models on the four studied datasets. Besides, the variant is more stable across datasets, unlike baselines that struggle on BGL. While KNN, DT, and MLP see a sharp performance drop on BGL (F1 $\leq$ 0.612), our method maintains a high F1-score (0.915), showing better robustness and consistency.


In the HDFS dataset, the grouping of log sequences by block ID likely results in more structured and homogeneous sequences, facilitating easier pattern recognition and classification for both the proposed Transformer-based method and the baseline models. 

In contrast, generating log sequences with varying lengths in the other datasets introduces additional challenges. When log sequences of varying lengths are transformed into \acs{MCV}s, variations in the volume of message counts across sequences may arise, presenting challenges in modeling. This variability in sequence lengths may pose difficulties for simpler machine learning methods like KNN, Decision Trees, and MLP, which may struggle to effectively learn and generalize patterns from sequences of different lengths across the entire dataset. \response{R2.4}{The performance outcomes may differ from previous empirical studies~\cite{he2016experience, wu2023effectiveness} due to variations in grouping settings, which lead to changes in log sequence lengths, the number of samples, and the anomaly ratio. These variations cause fluctuations in metric values, making direct comparisons of metrics across studies meaningless.}

The competitive and stable performance of our Transformer-based model in these datasets could be attributed to its inherent ability to handle variable-length sequences through mechanisms like self-attention, which allows the model to capture contextual information across sequences of different lengths effectively. The variability in baseline performances across datasets exhibits the sensitivity of simple machine learning methods to dataset characteristics and the varying lengths of log sequences.

\accepted{\heng{Elaborate the discussion: how do the results compare with the state-of-the-art in the literature; how the results are different/consistent across different datasets? It might be interesting to see how the performances are different/consistent for the different lengths of the input.}}

\begin{table*}[!ht]
    \centering
    \caption[Performance Comparison: Our Method vs. Baselines]{\response{R1.8}{Comparison of Performance: Our Model Variant (Random Embedding + Temporal Encoding (RTEE)) vs. Baseline Methods}}
    \label{two:table:supervised}

    \begin{tabular}{lcccc}
        \toprule
        \textbf{Dataset: HDFS} & \textbf{Prec} & \textbf{Rec} & \textbf{Spec} & \textbf{F1} \\ 
        \midrule
        Our method & 0.990 & 0.995 & 1.000 & 0.993 \\ 
        KNN & 0.992 & 0.996 & 1.000 & 0.994 \\ 
        DT  & 0.998 & 0.997 & 1.000 & 0.998 \\ 
        MLP & 0.998 & 0.997 & 1.000 & 0.998 \\ 
        \bottomrule
    \end{tabular}

    \vspace{1em}

    \begin{tabular}{lcccc}
        \toprule
        \textbf{Dataset: BGL} & \textbf{Prec} & \textbf{Rec} & \textbf{Spec} & \textbf{F1} \\ 
        \midrule
        Our method & 0.947 & 0.899 & 0.992 & 0.915 \\ 
        KNN & 0.376 & 0.867 & 0.925 & 0.524 \\ 
        DT  & 0.442 & 0.995 & 0.933 & 0.612 \\ 
        MLP & 0.449 & 0.961 & 0.933 & 0.612 \\ 
        \bottomrule
    \end{tabular}

    \vspace{1em}

    \begin{tabular}{lcccc}
        \toprule
        \textbf{Dataset: Spirit} & \textbf{Prec} & \textbf{Rec} & \textbf{Spec} & \textbf{F1} \\ 
        \midrule
        Our method & 0.973 & 0.930 & 0.998 & 0.951 \\ 
        KNN & 0.864 & 0.929 & 0.989 & 0.895 \\ 
        DT  & 0.966 & 0.996 & 0.997 & 0.981 \\ 
        MLP & 0.747 & 0.141 & 0.970 & 0.238 \\ 
        \bottomrule
    \end{tabular}

    \vspace{1em}

    \begin{tabular}{lcccc}
        \toprule
        \textbf{Dataset: Thunderbird} & \textbf{Prec} & \textbf{Rec} & \textbf{Spec} & \textbf{F1} \\ 
        \midrule
        Our method & 1.000 & 0.947 & 1.000 & 0.973 \\ 
        KNN & 0.931 & 0.990 & 0.994 & 0.959 \\ 
        DT  & 0.981 & 1.000 & 0.998 & 0.990 \\ 
        MLP & 0.855 & 0.443 & 0.987 & 0.584 \\ 
        \bottomrule
    \end{tabular}
\end{table*}

\paragraph{Experiment Set 2: Performance Comparison with NeuralLog}
\response{R1.8 \& R2.1 \& 3.3}{
Table~\ref{two:table:neurallog} presents the performance of our model variant incorporating Sentence Embedding and RTEE, as well as the performance of NeuralLog. On the one hand, these results validate the correctness of our method, showing that both models achieve comparable results with high Precision, Recall, and F1-scores across all datasets, and almost perfect Specificity ($\approx$1.000). On the other hand, the consistently strong performance of both methods suggests that the datasets may not be challenging enough to reveal significant differences between approaches.}

\begin{table*}[!ht]
    \centering
    \resizebox{\linewidth}{!}{
    \begin{threeparttable}
    \caption[Performance Comparison: Our Method vs. Baselines]{\response{R1.8\& R2.1 \&3.3}{Comparison of Performance: Our Model Variant (Sentence Embedding + Temporal Encoding) vs. NeuralLog}}
    \label{two:table:neurallog}
    \begin{tabular}{c|cc|cc|cc|cc}
    \hline
    \multirow{2}{*}{Metric} & \multicolumn{2}{c|}{HDFS} & \multicolumn{2}{c|}{BGL} & \multicolumn{2}{c|}{Spirit} & \multicolumn{2}{c}{Thunderbird} \\ \cline{2-9} 
                            & Our method & NeuralLog & Our method & NeuralLog & Our method & NeuralLog & Our method & NeuralLog \\ \hline
    Precision               & 0.989 & 0.992 & 0.987 & 0.986 & 0.987 & 1.000 & 0.999 & 1.000 \\
    Recall                  & 0.995 & 0.997 & 0.958 & 0.953 & 0.949 & 0.948 & 0.987 & 0.983 \\
    Specificity             & 1.000 & 1.000 & 0.999 & 0.999 & 1.000 & 1.000 & 1.000 & 1.000 \\
    F1                       & 0.992 & 0.995 & 0.972 & 0.969 & 0.968 & 0.973 & 0.993 & 0.992 \\
    \hline
    \end{tabular}
    \begin{tablenotes}
        \item[*] For BGL, Spirit, and Thunderbird, we use a fixed-window grouping configuration with a window size of 128 and a step size of 64.
    \end{tablenotes}
    \end{threeparttable}
    }
\end{table*}

\vspace{3mm}
\begin{answer*}{to RQ1}{}
The proposed Transformer-based anomaly detection model demonstrates competitive and stable performance when provided with log sequences of varying lengths, outperforming baseline models. In contrast, the simpler machine learning methods often struggle to adapt to the diverse lengths of log sequences, leading to unstable and inferior performance. \response{R1.8 \& R2.1 \& 3.3}{When compared with NeuralLog, another Transformer-based method, our approach achieves competitive results, thereby validating its effectiveness.}
\end{answer*}
\vspace{3mm}
\subsection{RQ2: How much does the sequential and temporal information within log sequences affect anomaly detection?}
\subsubsection{Objective}
A series of models (e.g., RNN, LSTM, Transformer) that are capable of catching the sequential or temporal information are employed for log-based anomaly detection. However, the significance of encoding these sequential or temporal information within log sequences is not clear, as there is no study that directly compares the performances with and without encoding the information. Moreover, recent studies show that anomaly detection models with log representations that ignore the sequential information can achieve competitive results compared with more sophisticated models~\cite{wu2023effectiveness}. In this research question, we aim to examine the role of sequential and temporal information in anomaly detection with our proposed Transformer-based model.

\subsubsection{Methodology}

In this research question, we compare the performance of our proposed model on different combinations of input features of log sequences. \accepted{\heng{double check the next two sentences}}We use the semantic embedding (as described in Sec~\ref{two:sec:parsing_embedding}) generated with log events in all the combinations. Additionally, we utilize three distinct inputs for the Transformer-based model: pure semantic embedding, the combination of semantic embedding and positional encoding, and the combination of\accepted{\heng{semantic embedding and?}} semantic embedding and temporal encoding. As the Transformer has no recurrent connections, it does not capture input token order like LSTM or GRU. It processes the whole input simultaneously with self-attention. Therefore, without positional and temporal encoding, the Transformer treats log sequences like bags of words (i.e., log events or messages), disregarding their order. This feature of the Transformer model enables us to evaluate the roles of positional and temporal information with the studied datasets in detecting anomalies. For temporal encoding, we explore two methods as outlined in Section~\ref{two:sec:enc}.


\subsubsection{Result}
Table~\ref{two:table:seq_temporal} shows the comparison of the performance when the model has different combinations of input features. From the results, we find that the model can achieve competitive performance when the input is the semantic embedding of log events in all the datasets. In this case, the sequential information within log sequences is ignored by the model. The log sequences are like a ``\textit{bag of log events}'' for the model. \response{R2.3}{With positional and temporal information, the model maintains consistent performance across datasets, suggesting a weak correlation between anomalies and sequential patterns or the sufficiency of semantic information for detection. However, incorporating Time2Vec for temporal encoding leads to a performance drop and slower convergence, as observed in the training curves. This may be due to the additional parameters introduced by Time2Vec, increasing the model's learning complexity for anomaly patterns.}

Among the variants with positional and temporal encodings, the one with positional encoding generally works better in all cases. This can be explained by the fact that positional encoding conveys simpler sequential information while temporal encoding carries richer information. However, the richer information may not be very helpful in detecting the anomalies within log sequences. The information redundancy further induces noises and increases learning difficulty for the model to learn the occurrence patterns associated with anomalies within log data. Moreover, among the two methods of temporal encoding, RTEE performs better than Time2Vec, likely due to Time2Vec's additional parameters, which increase the model's learning complexity.

The results reflect that the encoding of sequential and temporal information does not help the model achieve better performance in detecting anomalies in the four studied datasets. The occurrences of certain templates and semantic information within log sequences may be the most prominent indicators of anomalies.

\begin{table*}[!ht]
    \centering
    \caption[Anomaly Detection Performance Across Datasets]{\response{R2.3: Updated}{Comparison of the anomaly detection performance across varied positional and temporal encoding}}
    \label{two:table:seq_temporal}

    \begin{tabular}{lcccc}
        \toprule
        \textbf{Dataset: HDFS} & \textbf{Prec} & \textbf{Rec} & \textbf{Spec} & \textbf{F1} \\ 
        \midrule
        Semantic Only & 0.992 & 0.997 & 1.000 & 0.995 \\ 
        Semantic + Positional & 0.993 & 0.998 & 1.000 & 0.996 \\ 
        Semantic + RTEE & 0.989 & 0.995 & 1.000 & 0.992 \\ 
        Semantic + Time2Vec & 0.976 & 0.966 & 1.000 & 0.971 \\ 
        \bottomrule
    \end{tabular}

    \vspace{1em}

    \begin{tabular}{lcccc}
        \toprule
        \textbf{Dataset: BGL} & \textbf{Prec} & \textbf{Rec} & \textbf{Spec} & \textbf{F1} \\ 
        \midrule
        Semantic Only & 0.993 & 0.992 & 0.999 & 0.993 \\ 
        Semantic + Positional & 0.998 & 0.990 & 1.000 & 0.994 \\ 
        Semantic + RTEE & 0.986 & 0.982 & 0.998 & 0.984 \\ 
        Semantic + Time2Vec & 0.976 & 0.950 & 0.997 & 0.963 \\ 
        \bottomrule
    \end{tabular}

    \vspace{1em}

    \begin{tabular}{lcccc}
        \toprule
        \textbf{Dataset: Spirit} & \textbf{Prec} & \textbf{Rec} & \textbf{Spec} & \textbf{F1} \\ 
        \midrule
        Semantic Only & 0.926 & 0.962 & 0.999 & 0.971 \\ 
        Semantic + Positional & 0.988 & 0.964 & 0.999 & 0.976 \\ 
        Semantic + RTEE & 0.967 & 0.944 & 0.998 & 0.955 \\ 
        Semantic + Time2Vec & 0.955 & 0.887 & 0.997 & 0.920 \\ 
        \bottomrule
    \end{tabular}

    \vspace{1em}

    \begin{tabular}{lcccc}
        \toprule
        \textbf{Dataset: Thunderbird} & \textbf{Prec} & \textbf{Rec} & \textbf{Spec} & \textbf{F1} \\ 
        \midrule
        Semantic Only & 0.999 & 0.986 & 1.000 & 0.992 \\ 
        Semantic + Positional & 0.999 & 0.973 & 1.000 & 0.986 \\ 
        Semantic + RTEE & 0.997 & 0.979 & 1.000 & 0.988 \\ 
        Semantic + Time2Vec & 0.999 & 0.967 & 1.000 & 0.983 \\ 
        \bottomrule
    \end{tabular}
\end{table*}

\begin{answer*}{to RQ2}{}
Based on the studied datasets, integrating the sequential or temporal information generally does not contribute to the performance of anomaly detection on the studied datasets. On the contrary, encoding sequential or temporal information may increase learning complexity and negatively influence the model's performance.
\end{answer*}

\subsection{RQ3: How much do the different types of information individually contribute to anomaly detection?}
\subsubsection{Objective} From the RQ2, we find that the sequential and temporal information does not contribute to the performance of the anomaly detection model on the datasets. On the contrary, the encoding added to the semantic embedding may induce information loss for the Transformer-based model with the studied datasets. In this research question, our objective is to further analyze the different types of information associated with the anomalies in studied datasets. We aim to check how much semantic, sequential, and temporal information independently contributes to the identification of anomalies. This study may offer insight into the characteristics of anomalies within the studied datasets.

\subsubsection{Methodology}
In this research question, we utilize semantic embedding (generated with all-MiniLM-L6-v2~\cite{minilmmodel}) and the random embedding (See~\ref{sec:rq1:method}) to encode log events. By comparing the performance of models using these two input features, we can grasp the significance of semantic information in the anomaly detection task.


Moreover, we input only the temporal encoding to the model without the semantic embedding of the log events,\accepted{\heng{Need some explanation on how this is done in the Transformer structure}} with the aim of determining the degree to which anomalies can be detected solely by leveraging the temporal information provided by the timestamps of logs. We conduct the experiments with the two temporal encoding methods as in RQ2.

\subsubsection{Result}
Table \ref{two:table:semantic_temp} illustrates the results of the comparative analysis of anomaly detection performance using single input features across various datasets.

\paragraph{Semantic Embedding vs. Random Embedding} Utilizing semantic embedding as the input feature consistently yields high precision, recall, specificity, and F1 scores across all datasets. Conversely, employing random embedding results in slightly lower F1 scores, particularly noticeable on the BGL and Spirit datasets, underscoring the significance of semantic information in anomaly detection tasks. On the other hand, the models utilizing random embedding of log events as input also demonstrate strong discrimination capabilities regarding anomalies. This underscores the substantial contribution of occurrence information of log events alone to the tasks of anomaly detection. Such results validate the efficacy of employing straightforward vectorization methods (e.g., \acs{MCV}) for representing log sequences.

\paragraph{The role of temporal information} Analyzing the performance of models with temporal encodings as inputs across different datasets sheds light on the correlation between temporal information and anomalies in anomaly detection. Although there are variations in the numerical values, both temporal encoding methods consistently exhibit similar trends across datasets. Notably, the F1 scores of RTEE-only and Time2Vec-only models vary across datasets, indicating dataset-specific influences on anomaly detection. For instance, in the Spirit dataset, both RTEE-only and Time2Vec-only models exhibit extremely low F1 scores compared to other datasets, suggesting challenges in associating anomalies with the temporal patterns within the Spirit logs. In contrast, within the Thunderbird dataset, the temporal-only models demonstrate a discernible level of discriminatory capability towards anomalies, as evidenced by F1 scores of 0.670 and 0.640, indicating a more pronounced association between temporal data and annotated anomalies in this dataset.

From the results in RQ2, we find that the inclusion of temporal encoding to the input feature of the Transformer-based model may sometimes induce a performance loss. However, the experimental results in this RQ demonstrate that the temporal information displays some degree of discrimination, albeit to varying extents across different datasets. The findings suggest that semantic information and event occurrence information are more prominent than temporal information in anomaly detection. Most anomalies within public datasets may not be associated with temporal anomalies and can be easily identified with log events concerned. ~\response{R3.4}{Moreover, these findings generally align with previous empirical studies~\cite{landauer2024critical}, suggesting that our approach can be a valuable tool for analyzing new datasets to determine which types of information are most critical for anomaly detection.}

\begin{table*}[!ht]
    \centering
    \caption[Comparison of Anomaly Detection Performance Using Individual Input Features]{\response{R2.3: Updated}{Comparison of Anomaly Detection Performance Using Individual Input Features}}
    \label{two:table:semantic_temp}

    \begin{tabular}{lcccc}
        \toprule
        \textbf{HDFS (0.029)} & \textbf{Prec} & \textbf{Rec} & \textbf{Spec} & \textbf{F1} \\ 
        \midrule
        Semantic embedding & 0.992 & 0.997 & 1.000 & 0.995 \\ 
        Random embedding & 0.992 & 1.000 & 0.999 & 0.996 \\ 
        RTEE & 0.961 & 0.646 & 0.999 & 0.772 \\ 
        Time2Vec & 0.910 & 0.497 & 0.999 & 0.643 \\ 
        \bottomrule
    \end{tabular}

    \vspace{1em}

    \begin{tabular}{lcccc}
        \toprule
        \textbf{BGL (0.116)} & \textbf{Prec} & \textbf{Rec} & \textbf{Spec} & \textbf{F1} \\ 
        \midrule
        Semantic embedding & 0.993 & 0.992 & 0.999 & 0.993 \\ 
        Random embedding & 0.911 & 0.927 & 0.988 & 0.919 \\ 
        RTEE & 0.414 & 0.546 & 0.901 & 0.471 \\ 
        Time2Vec & 0.432 & 0.655 & 0.888 & 0.521 \\ 
        \bottomrule
    \end{tabular}

    \vspace{1em}

    \begin{tabular}{lcccc}
        \toprule
        \textbf{Spirit (0.070)} & \textbf{Prec} & \textbf{Rec} & \textbf{Spec} & \textbf{F1} \\ 
        \midrule
        Semantic embedding & 0.926 & 0.962 & 0.999 & 0.971 \\ 
        Random embedding & 0.999 & 0.767 & 1.000 & 0.868 \\ 
        RTEE & 0.112 & 0.694 & 0.570 & 0.192 \\ 
        Time2Vec & 0.110 & 0.686 & 0.573 & 0.190 \\ 
        \bottomrule
    \end{tabular}

    \vspace{1em}

    \begin{tabular}{lcccc}
        \toprule
        \textbf{Thunderbird (0.076)} & \textbf{Prec} & \textbf{Rec} & \textbf{Spec} & \textbf{F1} \\ 
        \midrule
        Semantic embedding & 0.999 & 0.986 & 1.000 & 0.992 \\ 
        Random embedding & 1.000 & 0.958 & 1.000 & 0.978 \\ 
        RTEE & 0.650 & 0.691 & 0.969 & 0.670 \\ 
        Time2Vec & 0.615 & 0.667 & 0.965 & 0.640 \\ 
        \bottomrule
    \end{tabular}

    \vspace{1em}

    \begin{tablenotes}
        \small
        \begin{minipage}{0.6\linewidth}
        \item[*] * The bracketed number after each dataset indicates the ratio of anomalies in the test sets. The F1 score of random guesses would tend to approach the proportion of the minority class (i.e., anomalies).
        \end{minipage}
    \end{tablenotes}

\end{table*}
\vspace{5mm}

\begin{answer*}{to RQ3}{}
While temporal information within log sequences may exhibit some degree of discrimination varying across datasets in anomaly detection, the semantic information and event occurrence information are more prominent for log-based anomaly detection based on the studied public datasets.~\response{R3.4}{Our approach can serve as a tool for identifying critical information for anomaly detection in new datasets.}
\end{answer*}

\vspace{3mm}

\section{Discussion}~\label{two:sec:discussion}
We discuss our lessons learned according to the experimental results.


\paragraph{Semantic information contributes to anomaly detection} The findings of this study confirm the efficacy of utilizing semantic information within log messages for log-based anomaly detection. Recent studies show classical machine learning models and simple log representation (vectorization) techniques can outperform complex \acs{DL} counterparts~\cite{wu2023effectiveness, yu2024deep}. In these simple approaches, log events within log data are substituted with event IDs or tokens, and semantic information is lost. However, according to our experimental results, the semantic information is valuable for subsequent models to distinguish anomalies, while the event occurrence information is also prominent.


\paragraph{We call for future contributions of new, high-quality datasets that can be combined with our flexible approach to evaluate the influence of different components in logs for anomaly detection.} The results of our study confirm the findings of recent works~\cite{landauer2024critical, yu2024deep}. Most anomalies may not be associated with sequential information within log sequences. The occurrence of certain log templates and the semantics within log templates contribute to the anomalies. This finding highlights the importance of employing new datasets to validate the recent designs of DL models (e.g., LSTM~\cite{du2017deeplog}, Transformer~\cite{le2021log}). Moreover, our flexible approach can be used off-the-shelf with the new datasets to evaluate the influences of different components and contribute to high-quality anomaly detection that leverages the full capacity of logs.

The publicly available log datasets that are well-annotated for anomaly detection are limited, which greatly hinders the evaluation and development of anomaly detection approaches that have practical impacts. Except for the HDFS dataset,\accepted{\heng{is the observation for the HDFS dataset different?}\xingfang{couldn't tell from the results}} whose anomaly annotations are session-based, the existing public datasets contain annotations for each log entry\accepted{\heng{entroy/message? keep wording consistent}} within log data, which implies the anomalies are only associated with certain specific log events or associated parameters within the events. Under this setting, the causality or sequential information that may imply anomalous behaviors is ignored.


\section{Threats to validity}~\label{two:sec:threats}

\accepted{\Foutse{this section could have been organized using the different types of threats to validity, i.e., construct, internal, reliability, etc ...this way we ensure to not miss some of them!}}

We have identified the following threats to the validity of our findings:

\accepted{\heng{Also discuss the generalizability issue introduced by the structure of the proposed model: a different structure may lead to different results.}\xingfang{it is included in the hyperparameter settings.}}

\paragraph{Construct Validity}

In our proposed anomaly detection method, we adopt the Drain parser to parse the log data. Although the Drain parser performs well and can generate relatively accurate parsing results, parsing errors still exist. The parsing error may influence the generation of log event embedding (i.e., logs from the same log event may have different embeddings) and thus influence the performance of the anomaly detection model. To mitigate this threat, we pass some extra regular expressions for each dataset to the parser. These regular expressions can help the parser filter some known dynamic areas in log messages and thus achieve more accurate results. \response{R2.1 \& R3.3}{Besides, we replicated the baseline method, but there may be discrepancies with the original work. The performance of our implementation is consistent with the results reported in the original paper under the same experimental conditions. To minimize this threat to validity, we strictly followed the original code and paper to ensure consistency, and our code is publicly available.} \response{R3.4}{Another potential threat to validity is noise in the datasets, including repetitive log sequences, errors, or inconsistencies~\cite{zhang2019robust, le2022log}. As our work did not address noise handling, this could affect the model's performance and our findings' generalizability. Future work may explore noise filtering or robust preprocessing techniques.}

\paragraph{Internal Validity}

There are various hyperparameters involved in our proposed anomaly detection model and experiment settings: 1) In the process of generating samples for both training and test sets, we define minimum and maximum lengths, along with step sizes, to generate log sequences of varying lengths. We have limited prior knowledge about the range of sequence length in which anomalies may reside. However, we set these parameters according to the common practices of previous studies, which adopt fixed-length grouping. 2) The Transformer-based anomaly detection model entails numerous hyperparameters, such as the number of Transformer layers, attention heads, and the size of the fully-connected layer. As the number of combinations is huge, we were not able to do a grid search. However, we referred to the settings of similar models and experimented with different combinations of hyperparameters, selecting the best-performing combination accordingly. \response{R1.6}{The choice of minimum and maximum sequence lengths, as well as the step size, are empirically determined based on prior works and our initial experiments~\cite{wu2023effectiveness}, as no standard values exist for these parameters. However, the results may still be influenced by the specific choice of these values, and alternative settings or datasets may yield different performance outcomes. To mitigate this threat, we apply these settings consistently across all experiments.}


\paragraph{External Validity}
In this study, we conducted experiments on four public log datasets for anomaly detection. Some findings and conclusions obtained from our experimental results are constrained to the studied datasets. However, the studied datasets are the most used ones to evaluate the log-based anomaly detection models. They have become the standard of the evaluation. As the annotation of the log datasets demands a lot of human effort, there are only a few publicly available datasets for log-based anomaly detection tasks. The studied datasets are representative, thus enabling the findings to illuminate prevalent challenges within the realm of anomaly detection. \response{R3.2}{The impact of different types of information within logs may vary across training schemes. Our conclusions are based on a supervised learning approach, and their validity under other paradigms remains uncertain. We experimented with a self-supervised learning scheme, but it did not yield satisfactory results and was therefore omitted in this study. Future work may explore alternative training methods to further investigate this aspect.}

\paragraph{Reliability}
The reliability of our findings may be influenced by the reproducibility of results, as variations in dataset preprocessing, hyperparameter tuning, and log parsing configurations across different implementations could lead to discrepancies. To mitigate this threat, we adhered to well-used preprocessing processes and hyperparameter settings, which are detailed in the paper. However, even minor differences in experimental setups or parser configurations may yield divergent outcomes, potentially impacting the consistency of the model's performance across independent studies.

\section{Conclusions}~\label{two:sec:conclusion}

\accepted{\heng{The text above is not necessary. Simply say that existing studies have used different types of information; however, it is not clear how the different types contribute...}}
The existing log-based anomaly detection approaches have used different types of information within log data. However, it remains unclear how these different types of information contribute to the identification of anomalies. In this study, we first propose a Transformer-based anomaly detection model, with which we conduct experiments with different input feature combinations to understand the role of different information in detecting anomalies within log sequences. The experimental results demonstrate that our proposed approach achieves competitive and more stable performance compared to simple machine learning models when handling log sequences of varying lengths. With the proposed model and the studied datasets, we find that sequential and temporal information do not contribute to the overall performance of anomaly detection when the event occurrence information is present\accepted{\heng{when the semantic information is present}}. The event occurrence information is the most prominent feature for identifying anomalies, while the inclusion of semantic information from log templates is helpful for anomaly detection models.\accepted{\heng{double check the two sentences: are they supported by RQ3? RQ3 only tells semantic embedding works the best, followed by random embedding (occurrence of logs)}} Our results and findings generally confirm that of the recent empirical studies and indicate the deficiency of using the existing public datasets to evaluate anomaly detection methods, especially the deep learning models. Our work highlights the need to utilize new datasets that contain different types of anomalies and align more closely with real-world systems to evaluate anomaly detection models. Our flexible approach can be readily applied with the new datasets to evaluate the influences of different components and enhance anomaly detection by leveraging the full capacity of log information.

\bmhead{Supplementary information}

The source code of the proposed method is publicly available in our supplementary material package~\footnote{\href{https://github.com/mooselab/suppmaterial-CfgTransAnomalyDetector}{https://github.com/mooselab/suppmaterial-CfgTransAnomalyDetector}}.



\bmhead{Acknowledgements}
We sincerely appreciate the funding support from the Natural Sciences and Engineering Research Council of Canada (NSERC, RGPIN-2021-03900) and the Fonds de recherche du Québec – Nature et technologies (FRQNT, 326866) for their funding support for this work. We also acknowledge the computational resources provided by Calcul Québec (calculquebec.ca) and the Digital Research Alliance of Canada (alliancecan.ca), which helped make this research possible.






\noindent



\bibliography{main}

\end{document}